\documentclass
[aps,prl,twocolumn,showpacs,floats,amsmath]{revtex4}%
\usepackage{graphicx}
\usepackage{amsmath}
\usepackage{amsfonts}
\usepackage{amssymb}%
\usepackage{epsf}

\newcommand{\be}{\begin{equation}}
\newcommand{\ee}{  \end{equation}}
\newcommand{\ba}{\begin{eqnarray}}
\newcommand{\ea}{  \end{eqnarray}}

\begin{document}

\title{Gaussian Decoherence and Gaussian Echo from Spin Environments} 
\author{W.H. Zurek}
\author{F.M. Cucchietti}
\author{J.P. Paz} 
\affiliation{Theoretical Division, MS B213, Los Alamos National 
Laboratory, Los Alamos, NM 87545} 
\date{\today} 
 
\begin{abstract}
We examine an exactly solvable model of decoherence -- a spin-system interacting with a collection 
of environment spins. We show that in this simple model (introduced some time ago to illustrate 
environment--induced superselection) generic assumptions about the coupling strengths typically lead 
to a non-Markovian (Gaussian) suppression of coherence between pointer states. We explore the 
regime of validity of this result and discuss its relation to spectral features of the environment. We also 
consider its relevance to Loschmidt echo experiments (which measure, in effect, the fidelity between 
the initial state and the state first evolved forward with a Hamiltonian ${\cal H}$, and then ``unevolved''
with (approximately) $-{\cal H}$). In particular, we show that for partial reversals (e.g., when only a 
part of the total Hamiltonian changes sign) fidelity may exhibit a Gaussian dependence on the time of reversal that is independent of the details of the reversal procedure: It just depends on what part of the Hamiltonian gets ``flipped'' by the reversal. This puzzling behavior was observed in several NMR experiments. Natural candidates for such two environments (one of which is easily reversed, while the other is ``irreversible'') are suggested for the experiment involving ferrocene.

\end{abstract} 

\pacs{03.65.Yz;~03.67.-a}

\maketitle 

\section{Introduction}

``It seems very important to us... that the idea and genesis of randomness can be made
rigorously precise also if one rigorously follows the determinism; the law of large numbers
comes then not as a mystical principle and not as a purely empirical fact, but as a simple 
mathematical result..." wrote Marian Smoluchowski in his posthumously published paper
\cite{Smol1918}. At that time determinism meant {\it classical} determinism -- the underlying 
equations of motion that determined a trajectory of a classical particle. One could,
however, develop simple stochastic models that encapsulated effects of that exact dynamics.
That was the essence of the approach that led to the Smoluchowski equation (which is still
widely used today, a century after it was derived using this strategy). 

Quantum theory forces one to reassess the relation between determinism and randomness:
Chance plays a different role in the quantum domain. According to Bohr and Born, quantum  
randomness is fundamental: A measurement on a quantum system -- according to the Copenhagen interpretation -- necessarily involves a classical apparatus. The outcome of the measurement
is randomly selected with probability given by the famous rule that connects probability to 
amplitude ($p_k=|\psi_k|^2$) conjectured by Max Born. The Copenhagen view of the quantum 
Universe was challenged by Everett, who half a century ago noted that it is possible to imagine 
that our Universe is all quantum, and that its global evolution is deterministic. Randomness 
will appear only as a result of the local nature of subsystems (such as an apparatus or an observer)
\cite{Zurek82,Zurek03}.

It is not our aim here to recapitulate this well known story, except to point out that it sheds a rather
different light on the relation between determinism and randomness than did classical physics.
The key insight of Smoluchowski contained in the quote above is, however, still correct -- perhaps
even more deeply correct -- in the quantum setting. 
Entanglement, the quintessential quantum phenomenon,
which plays such an important role in the approach of Everett is central for its
validity. We illustrate here only one aspect of these connections -- decoherence which is caused 
by entangling interactions between the system and the environment. 

The story that will unfold makes one more connection with Smoluchowski: It touches on the debate
about the origins and nature of irreversibility between his two former professors -- Boltzmann 
and Loschmidt -- as the evolution responsible for the buildup of correlations which lead 
to decoherence can be (approximately) reversed in suitable settings, allowing for the study of  
``Loschmidt echo'' \cite{horaciobook,horacioexponential,horaciosaturation,horaciodifusion}. 

\section{Spin Decoherence Model}

A single spin--system ${\cal S}$ (with states $\left\{ \left|0\right>, \left|1\right> \right\}$)
interacting with an environment ${\cal E}$ of many independent spins
($\left\{ \left| \uparrow_k \right>, \left| \downarrow_k \right> \right\}$,
$k=1..N$) through
the Hamiltonian
\be
{\cal H_{SE}} =\left(\left|0\right>\left<0\right| - 
\left|1\right>\left<1\right| \right) \sum_{k=1}^N \frac{g_k}{2}
\left( \left| \uparrow_k \right> \left< \uparrow_k \right| - 
\left|\downarrow_k \right>\left<\downarrow_k \right| \right)
\label{hamiltonian0}
\ee
may be the simplest solvable model of decoherence. 
It was introduced
some time ago \cite{Zurek82} to show that relatively 
straightforward
assumptions about the dynamics can lead to the emergence of a preferred
set of pointer states due to einselection (environment--induced superselection) 
\cite{Zurek82,deco}.
Such models have gained additional importance in the past decade because of
their relevance to quantum information processing \cite{spindeco}. 

The purpose of our paper is to show that -- with a few additional natural and simple
assumptions -- one can evaluate the exact time dependence of the reduced density matrix, and
demonstrate that the off--diagonal components display a Gaussian 
(rather than exponential) decay \cite{ZCP}.
In effect, we exhibit a simple soluble example of a situation where the usual
Markovian \cite{Kossakowski} assumptions about the evolution of a quantum open
system are not satisfied.
Apart from their implications for decoherence, our results are also relevant to quantum error
correction \cite{ErrorCorrection} where precise precise knowledge of the 
dynamics is essential to select an efficient strategy. Moreover, while the model Hamiltonian
of Eq.~(\ref{hamiltonian0}) is very specific, it suggests generalizations that lead one to conclude 
that Gaussian decay of polarization may be common, and specify when a reversal of 
the Hamiltonian evolution in a part of the spin environment naturally leads to a Gaussian 
dependence of the return signal on the time of reversal, a feature of Loschmidt echo observed in NMR experiments.

To demonstrate the Gaussian time dependence of decoherence we first write down a
general solution for the model given by Eq.~(\ref{hamiltonian0}). 
Starting with:
\be
\left| \Psi_{\cal SE}(0)\right> = 
(a \left|0\right>+b \left|1\right>) \bigotimes_{k=1}^N
\left( \alpha_k \left| \uparrow_k \right> + 
\beta_k \left|\downarrow_k \right> \right),
\label{initialstate}
\ee
the state of ${\cal SE}$ at an arbitrary time is given by:
\be
\left| \Psi_{\cal SE}(t)\right> = 
a \left|0\right> \left|{\cal E}_0 (t)\right> 
+b \left|1\right>  \left|{\cal E}_1 (t)\right>
\label{phit}
\ee
where
\ba 
\left|{\cal E}_0 (t)\right> & = & \bigotimes_{k=1}^N
\left( \alpha_k e^{i g_k t/2} \left| \uparrow_k \right> 
+ \beta_k e^{-i g_k t/2} \left|\downarrow_k \right> \right)  \nonumber \\ 
&=& \left|{\cal E}_1 (-t)\right>.
\label{environ}
\ea
The reduced density matrix of the system is then:
\ba
\rho_{\cal S} & = & {\rm Tr} _{\cal E} \left| \Psi_{\cal SE}(t)\right> 
\left< \Psi_{\cal SE}(t)\right| \nonumber \\
& = & |a|^2 \left|0\right>\left<0\right|+ a b^{*} r(t)
\left|0\right>\left<1\right| \nonumber \\ 
& + & a^{*} b r^{*}(t) \left|1\right>\left<0\right| + |b|^2 \left| 1 \right> \left< 1 \right|,
\label{reducedrho}
\ea
where the {\it decoherence factor}
$r(t)=\left<{\cal E}_1 (t)|{\cal E}_0 (t)\right>$
can be readily obtained:
\ba
r(t)&=&\prod_{k=1}^N 
\left( |\alpha_k|^2 e^{i g_k t} + |\beta_k|^2 e^{-i g_k t} \right).
\label{roft}
\ea

It is straightforward to see that $r(0)=1$, and that for $t>0$ it will decay
to zero, so that the typical fluctuations of the off-diagonal terms of
$\rho_{\cal S}$ will be small for large environments, since:
\be
\left<|r(t)|^2 \right>=2^{-N} \prod_{k=1}^N \left( 1+(|\alpha_k|^2 - 
|\beta_k|^2)^2 \right),
\label{rsoft}
\ee
Here $\left<...\right>$ denotes a long time average \cite{Zurek82}.
Clearly, 
$\left<|r(t)|^2 \right> \underset{N\rightarrow \infty}{\longrightarrow} 0$,
leaving $\rho_{\cal S}$ approximately diagonal in a mixture of the pointer states 
$\left\{ \left|0\right>, \left|1\right> \right\}$ which retain preexisting
classical correlations.

This much was known since \cite{Zurek82}. The aim of this paper is to show
that, for a fairly generic set of assumptions, the form of $r(t)$ can be
further evaluated and that -- quite universally -- it turns out to be 
approximately Gaussian in time. Thus, the simple model of Ref. \onlinecite{Zurek82}
predicts a universal (Gaussian) form of the loss of quantum coherence, whenever
the couplings $g_k$ of Eq.~(\ref{hamiltonian0}) are sufficiently concentrated near
their average value so that their standard deviation
$\left<(g_k-\left<g_k\right>)^2\right>$ exists and is finite. When this condition
is not fulfilled other sorts of time dependence become possible. In particular,
$r(t)$ may be exponential when the distribution of couplings is a Lorentzian.

We shall also consider implications of the predicted time dependence of $r(t)$ for echo experiments. 
In particular, the group of Levstein and Pastawski \cite{horaciobook,horacioexponential,horaciodifusion,horaciosaturation},
have carried out experiments that aim to implement 
time reversal of dynamics, as was suggested long time ago by Loschmidt \cite{loschmidt}, who
used time reversal as a counterargument to Boltzmann's ideas about H-theorem and the origins of
irreversibility. Boltzmann's reported (possibly apocryphal) reply ``Go ahead and do it!", 
which may reflect his belief in the molecular disorder hypothesis \cite{kuhn}, points to the
origin of the difficulty in implementing such reversal in practice for {\it all} of the relevant 
degrees of freedom. It is nevertheless possible in some settings to carry out ``Loschmidt echo experiments'' that approximate Loschmidt's original idea \cite{loschmidt}. 

When the reversal is successful for only some of the relevant degrees of freedom (${\cal E}''$) but 
does not encompass all of the environment ${\cal E}$ (leaving behind ``unreversed'' ${\cal E}'$) 
the result is a {\it partial Loschmidt echo} (also dubbed ``Boltzmann echo'' \cite{jacquod}). As in Ref. 
\cite{LEdeco}, we interpret the decay in the Loschmidt echo as the effect of coupling to a second environment. We shall study the partial Loschmidt echo in the context of the simple model of Eq.~(\ref{hamiltonian0}) 
and Ref. \cite{Zurek82}, and conclude that its basic implications may generalize to a much broader 
range of dynamics relevant to NMR experiments. 

In our case the state of all the degrees of freedom after a partial reversal (that happens at $t=t_R$)
is given by:
\be
|\Phi_{{\cal S}{\cal E}' {\cal E}''}(t)\rangle=  e^{i H_{{\cal S}{\cal E}'}t}  e^{i H_{{\cal S}{\cal E}''}(2t_R-t)}|\Phi_{{\cal S}{\cal E}' {\cal E}''}(0)\rangle
\label{echo1}
\ee
The echo signal measured in the experiments concerns only a part of the whole -- the system 
${\cal S}$. It is given by:
$$\mu(2t_R)= Tr \rho_{\cal S}(t=0)\rho_{\cal S}(t=2t_R) \ . $$
This is in effect the fidelity of the state of ${\cal S}$. We shall express $\mu(t)$ in terms of decoherence factors corresponding to ${\cal E}'$ and ${\cal E}''$: This follows from a straightforward generalization 
of Eqs.~(\ref{reducedrho},~\ref{roft}) to the case of partial Loschmidt echo with two environments, only one of which 
gets reversed. 

\section{Gaussian Decoherence}

Evaluating time dependence of the decoherence factors for ${\cal E}'$ and ${\cal E}''$ 
is therefore our first goal. To this end we carry out multiplication of Eq.(\ref{roft}), 
re--expressing $r(t)$ as a sum:
\ba
r(t) &=& \prod_{k=1}^N |\alpha_k|^2  e^{i t \sum_{n} g_n} 
+ \sum_{l=1}^N |\beta_l|^2 \prod_{k\ne l}^N |\alpha_k|^2 \times \nonumber \\
& & e^{i t(-g_l +\sum_{n\ne l} g_n )} 
+ \sum_{l=1}^N\sum_{m\ne l}^N |\beta_l|^2 |\beta_m|^2 \times \nonumber \\
& &\prod_{k\ne l,m}^N |\alpha_k|^2 
e^{\left[i t (-g_l-g_m + \sum_{n\ne l,m}^N g_n )\right]} 
+...
\label{rexpansion}
\ea
Decoherence factor is then a sum of $2^N$ complex contributions with fixed absolute values
and with phases that rotate at the rate given by the eigenvalues of the total Hamiltonian. 

\begin{figure}
\centering \leavevmode
\epsfxsize 3.2in
\epsfbox{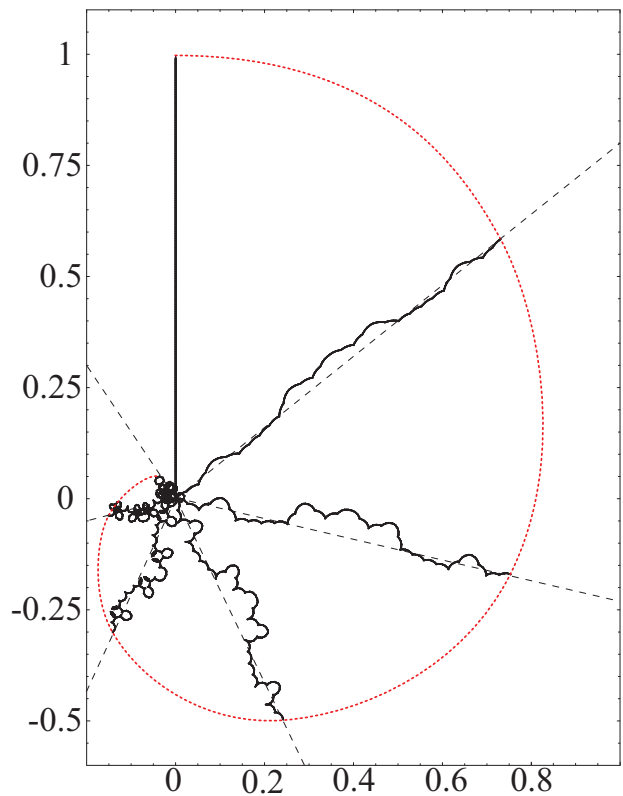}
\caption{Decoherence factor $r(t)$ decomposed as a sum of complex terms as in Eq. (9), for
$|\alpha_k|^2=|\beta_k|^2=1/2$ and $N=8$ spins in the environment with random couplings
$g_k$ from a uniform distribution. 
The times plotted are $t=0$, (at ``noon''), and $t=0.25$, $0.5$, $0.75$, $1$, $1.25$, and $1.5$
in a ``clockwise direction'': The coordinate of the complex plane is rotated clockwise by an angle $2\pi /7$ for each $r(t)$ (dashed lines), starting with a vertical axis for $t=0$. Notice the random walk-like behavior of $r(t)$. The dotted curve line is the envelope of the random walks -- the net decoherence factor decaying with a Gaussian form in accord with our discussion. }
\label{clock} 
\end{figure}

Decay of $r(t)$ can be understood (see \cite{Zurek82}) as a progressive randomization of a walk in a complex plane: 
At $t=0$ all of the phases are the same so all of the steps -- all of the contributions to the decoherence 
factor -- add up in phase yielding $|r(t=0)|=1$. However, as time goes on, these phases rotate at various rates so $r(t)$ is a terminal point of what becomes in time a random walk (on a complex plane) where 
the directions of various steps are uncoordinated (see Fig. \ref{clock}). 

This view of the decay of $r(t)$ is the first instance where the random walk analogy is useful in our paper. 
The terminal point of this random walk determines decoherence factor. Another random walk
-- this time in energy -- can be invoked in computing eigenvalues of the total Hamiltonian. These 
eigenvalues are responsible for the rotation rates of the individual steps that contribute to $r(t)$. 
We shall now see that this random walk in energies is responsible for the (typically Gaussian) 
decay of the decoherence factor. 

To exhibit the Gaussian nature of $r(t)$ we start by noting that there
are $\binom{N}{0}$, $\binom{N}{1}$, $\binom{N}{2}$, ... etc. terms in the
consecutive sums above. 
The binomial pattern is clear, and can be made even more apparent by assuming 
that $\alpha_k=\alpha$ and $g_k=g$ for all $k$. Then,
\be
r(t)= \sum_{l=0}^N \binom{N}{l} |\alpha|^{2(N-l)} |\beta|^{2l} e^{ig(N-2l)t},
\label{binomial}
\ee
i.e., $r(t)$ is the binomial expansion of $r(t)=\left(|\alpha|^2 e^{igt}+
|\beta|^2 e^{-igt}\right)^N$. 

We now note that, as follows from the Laplace-de Moivre theorem \cite{Gnedenko}, 
for sufficiently large $N$ the coefficients of the binomial expansion
of Eq.~(\ref{binomial}) can be approximated by a Gaussian:
\ba
\binom{N}{l} |\alpha|^{2(N-l)} |\beta|^{2l} \simeq 
\frac{\exp{\left[-\frac{(l-N|\beta|^2)^2}{2 N |\alpha \beta|^2}\right]}}
{\sqrt{2 \pi N |\alpha \beta|^2}} .
\label{gaussian}
\ea
This limiting form of the distribution of the eigenenergies 
of the composite ${\cal SE}$ system
immediately yields our main result:
\be
|r(t)| = \exp(-2N|\alpha \beta|^2 (gt)^2)
\label{dfact}
\ee
So, $r(t)$ is approximately Gaussian since it is a Fourier transform of an approximately Gaussian 
distribution of the eigenenergies of the total Hamiltonian resulting from all the possible combinations 
of the couplings with the environment. 

A few quick comments on the above form of the decoherence factor may be in order: We note
that in the limit of large $Ng^2$ it predicts ``instantaneous'' decay of quantum coherence.
We also note that when $\alpha\beta=0$ the environment is incapable of decohering the system
(as it is then in an eigenstate of the global Hamiltonian, so the ``measurement-like evolution'' that is
at the heart of decoherence is impossible). Last but not least, we note that when the environment 
is mixed, decoherence will proceed unimpeded, and that it will be most efficient when the mixture is
perfect -- when $|\alpha|^2=|\beta|^2=\frac 1 2$. 

\section{Law of Large Numbers and Energies}

\begin{figure}
\centering \leavevmode
\epsfxsize 3.2in
\epsfbox{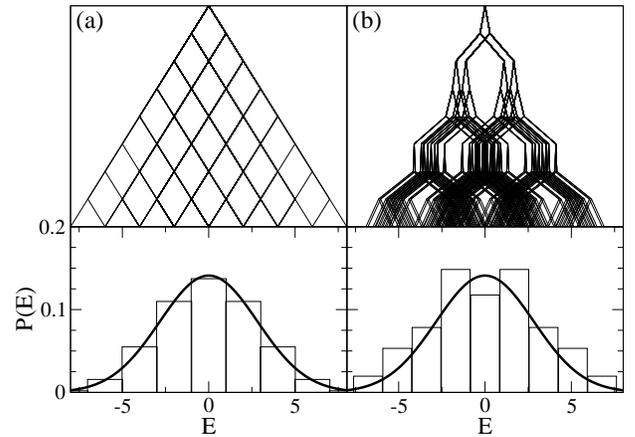}
\caption{The distribution of the energies obtains from the random walks with the
steps given by the coupling size and in the direction ($+g_k$ or $-g_k$) biased
by the probabilities $|\alpha_k|^2$ and $|\beta_k|^2$ as in Eq.~(\ref{Ldosrw})
(although in these examples we set $|\alpha_k|^2=1/2$).
(a) When all the couplings
have the same size $g_k=g$ (Eq.~(\ref{binomial})), 
a simple Newton's triangle leads to an approximate
Gaussian for the distribution of energies. (b) When the couplings
randomly differ from step to step (Eq.~(\ref{rexpansion})), 
the resulting distribution still
approaches an approximately Gaussian envelope for large $N$. }
\label{Figure1} 
\end{figure}

To yield a Gaussian decay of $|r(t)|$, the set of all the resulting eigenenergies of the total Hamiltonian must have 
an (approximately) Gaussian distribution. This behavior is generic, a result of the law of large numbers \cite{Gnedenko}: these energies can be thought of as the terminal points of an $N$--step random walk. The contribution of the $k$--th spin of the environment to the random energy is $+g$ or
$-g$ with probability $|\alpha|^2$ or $|\beta|^2$ respectively (Fig. \ref{Figure1}-a).

The same argument can be carried out in the more general case of Eq.~(\ref{rexpansion}).
The ``random walk'' picture that yielded the distribution of the couplings
remains valid (see Fig. \ref{Figure1}-b).
However, now the individual steps in the random walk are not all equal. Rather,
they are given by the set $\left\{g_k \right\}$ (see Eq.~\ref{hamiltonian0}) 
with each step $g_k$ taken just once in a given walk.
There are $2^N$ such distinct random walks. This exponential proliferation of the contributing
coupling energies allows one to anticipate rapid convergence to the universal Gaussian
form of their distribution, and, therefore, of the decoherence factor $r(t)$. 

Indeed, we can regard eigenenergies resulting from the sums of $g_k$'s as a
random variables. Its probability distribution is given by products of the
corresponding weights. 
That is, the typical term in Eq.~(\ref{rexpansion}) is of the form:
\be
p_W e^{iE_W t} \equiv \left( \prod_{k\in W^+} |\alpha_k|^2e^{i g_kt}\right) 
 \left(\prod_{k\in W^-}|\beta_k|^2e^{-i g_kt} \right).
\ee
The resulting terminal energy is
\be
E_W=\sum_{k\in W^+} g_k-\sum_{k\in W^-} g_k,
\ee
and the cumulative weight $p_W$ is given by the corresponding product of
$|\alpha_k|^2$ and $|\beta_k|^2$. Each such specific random walk $W$
corresponding to a given combination of right ($k\in W^+$) and left ($k\in W^-$)
``steps" (see Figs. \ref{clock} and \ref{Figure1}) contributes to the distribution of energies only once. 
The terminal points $E_{W}$ may or may not be degenerate: 
As seen in Fig. \ref{Figure1}, in the degenerate case, the whole
collection of $2^N$ random walks ``collapses" into $N+1$ terminal
energies. More typically, in the non-degenerate case (also displayed in Fig. \ref{Figure1}),
there are $2^N$ different terminal energies $E_W$. 
In both cases, the ``envelope" of the distribution $P(E_W)$ should be Gaussian, 
as we shall show below.

In contrast to the usual classical random walk scenario (where each event corresponds to 
specific random walk) in this quantum setting {\it all} of the random walks in the ensemble
contribute simultaneously -- evolution happens because the system is in a superposition of its 
energy eigenstates. The resulting decoherence factor $r(t)$ can be viewed as the characteristic 
function \cite{Gnedenko} (i.e., the Fourier transform) of the distribution of eigenenergies $E_W$. Thus,
\be
r(t)=\int e^{iEt} \eta(E) dE,
\label{rLDOS}
\ee
where the strength function $\eta(E)$, also known as the local density of states (LDOS) 
\cite{LDOS} is defined in general as
\be
\eta(E)=\sum_\lambda |\left< \Psi_{\cal SE}(0)| \phi_\lambda \right>|^2 \delta(E-E_\lambda).
\ee
Above $\left|\phi_\lambda\right>$ are the eigenstates of the full Hamiltonian
and $E_\lambda$ its eigenenergies. In our particular model (Eq.~\ref{hamiltonian0}) 
the eigenstates are associated with all possible random walks in the set $W$, 
and therefore 
\be
\eta(E)=\sum_W p_W \delta(E-E_W).
\label{Ldosrw}
\ee
Decoherence in our model is thus directly related to the characteristic function of the distribution of 
eigenenergies $\eta(E)$.
Moreover, since the $E_W$'s are sums of $g_k$'s, $r(t)$ is itself a product of characteristic functions 
of the distributions of the couplings $\{g_k\}$, as we have already seen in the example of
Eq. (\ref{roft}). Thus, the distribution of $E_W$ belongs to the
class of the so--called {\it infinitely divisible distributions} \cite{Gnedenko,breiman}. 

\begin{figure}
\centering 
\epsfxsize 3.2in
\epsfbox{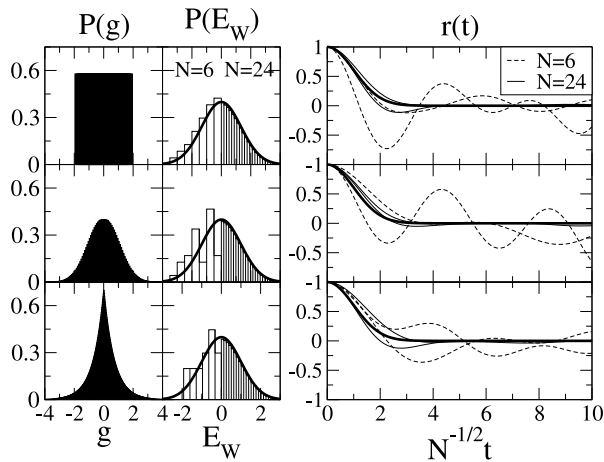}
\caption{(Left panels) Assumed distribution of the couplings $g_k$, from top to bottom: uniform, Gaussian,
and exponential. 
(Center panels) Resulting distribution of the
eigenenergies $E_W$ (center panels) for $N=6$ ($E_W<0$) and $N=24$
($E_W>0$). In the case of $|\alpha_k|^2=1/2$ this distribution is in effect the
``strength function" (local density of states). (Right panels) 
Decoherence factor $r(t)$ for different initial conditions with $N=6$ (dashed lines), 
$N=24$ (thin solid lines) and the average (bold line) rapidly approaches a Gaussian 
whenever couplings have a finite variance.}
\label{Figure2} 
\end{figure}

The behavior of the decoherence factor $r(t)$ 
-- characteristic function of an infinitely divisible distribution --
depends only on the average and variance of the distributions of couplings
weighted by the initial state of the environment
\cite{Gnedenko,breiman}.
The remaining task is to calculate $\eta(E)$, which can be obtained through the
statistical analysis of the random walk picture described above. If we
denote $x_k$ the random variable that takes the value $+g_k$ or $-g_k$ with probability
$|\alpha_k|^2$ or $|\beta_k|^2$ respectively, then its mean value
$a_k$ and its variance $b_k$ are
\ba
a_k&=&(|\alpha_k|^2-|\beta_k|^2)g_k, \nonumber \\
b_k^2&=&g_k^2-a_k^2=4|\alpha_k|^2|\beta_k|^2g_k^2.
\ea
The behavior of the sums of $N$ random variables $x_k$ (and thus, of their characteristic
function) depends on whether the so--called Lindeberg condition holds
\cite{Gnedenko}. It is expressed in terms of the cumulative variances
$B_N^2=\sum b_k^2$, and it is satisfied when the probability of the large
individual steps is small; e.g.:
\be
P(\underset{1\le k \le N}{{\rm max}} |g_k-a_k| \ge \tau B_N)
\underset{N\rightarrow\infty}{\longrightarrow} 0,
\ee
for any positive constant $\tau$. In effect, Lindeberg condition demands that the variance of couplings
exist and be finite -- i.e., that $B_N$ be
finite: when it is met, the resulting distribution
of energies $E=\sum x_k$ is Gaussian
\be
P\left( \frac{E-{\overline E}_N}{B_N}<x\right) 
\underset{N\rightarrow\infty}{\longrightarrow}
\int_{-\infty}^{x} e^{-s^2/2}ds,
\ee
where ${\overline E}_N=\sum_k a_k$. In terms of the LDOS this implies
\be
\eta(E)\simeq \frac{1}{\sqrt{2\pi B_N^2}}
\exp{\left(\frac{-(E-{\overline E}_N)^2}{2 B_N^2}\right)},
\ee
an expression in excellent agreement with numerical results already for modest
values of $N$.
This distribution of energies yields a corresponding approximately
Gaussian time--dependence of $r(t)$, as seen in Fig. \ref{Figure2}. Moreover, at least
for short times of interest for, say, quantum error correction, $r(t)$ is
approximately Gaussian already for relatively small values of $N$. This
conclussion holds whenever the initial distribution of the couplings has a
finite variance. The general form of $r(t)$ after applying the Fourier transform
of Eq. (\ref{rLDOS}) is
\be
r(t)\simeq e^{i {\overline E}_N t} e^{-B_N^2 t^2/2}.
\ee

\begin{figure}
\centering \leavevmode
\epsfxsize 3.2in
\epsfbox{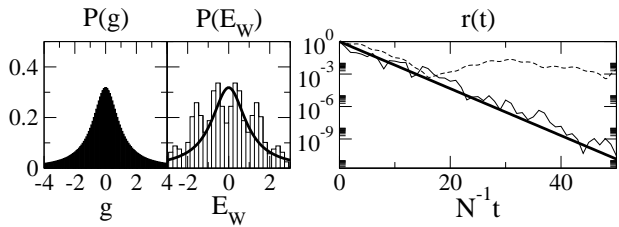}
\caption{Same as Fig. 3 but for a Lorentzian distribution of the couplings
${g_k}$. In this case $r(t)$ decays exponentially. 
The histogram and the dashed line in $r(t)$ correspond to $N=20$,
the straight thin line is a particular case for $N=100$ and the thick line is 
the average. We note that the convergence is slower than in the Gaussian case 
of Fig. 2, because realizations of ${g_k}$ are more likely to have 
one or two dominant couplings. Therefore, although the average shows a clear
exponential decay, fluctuations are noticeable even for large $N$.
Notice also that the logarithmic scale confirms the long time saturation of
$r(t)$ at $\sim 2^{-N/2}$, Eq~(\ref{rsoft}).}
\label{Figure3} 
\end{figure}

It is also interesting to investigate cases when Lindeberg condition is not
met. Here, the possible limit distributions are given by the stable (or L\'{e}vy)
laws \cite{breiman}.
One interesting case
that yields an exponential decay of the decoherence factor is the
Lorentzian distribution of couplings (see Fig. \ref{Figure3}). It 
can be expected e.g. in the effective dipolar couplings to
a central spin in a crystal \cite{Abragam}.
Further intriguing questions concern the robustness of our conclusion under the
changes of the model. We have begun to address this issue elsewhere
\cite{CPZ} but, for the time being, we only note that the addition
of a strong self--Hamiltonian proportional to $\sigma_x$ changes the nature of the time
decay \cite{Dobrovitski,CPZ}. On the other hand, small changes of the environment
Hamiltonians, like for instance truncated dipolar interactions,
\begin{equation}
{\cal H_E}=\sum_{i,j} g_{ij} \left( 2 \sigma_i^z \sigma_j^z - 
\sigma_i^x \sigma_j^x - \sigma_i^y \sigma_j^y \right),
\end{equation}
seem to preserve the Gaussian nature of $r(t)$ \cite{CPZ}. This universality of Gaussian decoherence
extends beyond the short-time regime where it was emphasized in Ref. \onlinecite{BHS}. It arises as 
a consequence of the central limit theorem that leads to Gaussian distribution of the eigenenergies, a limiting behavior that can be expected for reasons pointed out above (see also \cite{ZCP}) under generic conditions in many body systems \cite{HMH}.

\section{Partial Reversal and Gaussian Echo}

Let us now consider a Loschmidt echo -- reversal of the sign of the Hamiltonian -- carried out at 
a time $t_R$. In our model it can be implemented by appropriate ``flipping'' of the spins in 
the environment. We first note that the measured observable $\mu(t)$ signal can be readily 
related to the decoherence factor:
\ba 
\mu(t) & = & Tr \rho_{\cal S}(t)\rho_{\cal S}(0) \nonumber \\
& = & (a^*\langle0| + b^*\langle1|)
\rho_{\cal S}(t)(a|0\rangle + b|1\rangle) \nonumber \\
& =  & |a|^4|+ |b|^4 + 2 |ab|^2 Re ~ r(t)
\label{echo2}
\ea
For a complete Loschmidt echo, the sign of the the whole Hamiltonian would be reversed at $t=t_R$, 
so for $t > t_R$;
\be
\mu(t)= |a|^4| + |b|^4 + 2 |ab|^2 Re ~ r(t_R-(t-t_R)) \ . 
\label{echo3}
\ee
Hence, the decoherence factor is now $r(2t_R-t)$, and the system will return to its initial state at $t=2t_R$.

We now suppose with Petitjean and Jacquod \cite{jacquod} that only a part of the Hamiltonian is reversed (e.g., only some of the spins --  spins in ${\cal E}''$ -- get flipped). 
In our model, environments ${\cal E}'$ and ${\cal E}''$ do not interact. Thus, the net decoherence 
factor is a product of the decoherence factors coming from each environment,
\be
r(t)=r'(t) r''(t),
\ee
with
\ba
r'(t) & \approx & e^{i E_{N'}t} \exp \left( -B_{N'}^2 t^2/2 \right) \nonumber \\
& = & e^{i \sum g_k' t} \exp \left( -\sum (g_k')^2 t^2/2 \right)
\label{ffor}
\ea
and
\ba
r''(t) & \approx & e^{i  {E_{N''}(2t_R-t)} } \exp \left( -B_{N''}^2 (2t_R-t)^2/2 \right) \nonumber \\ 
& = & e^{i \sum g_k''(2t_R-t)} \exp \left( - \sum (g_k'')^2 (2t_R-t)^2/2 \right) \ .
\label{frev}
\ea
Since the time reversal only applies to ${\cal E}''$, 
\be
\mu(t) = |a|^4| + |b|^4 + 2 |ab|^2 Re ~ r'(t) r''(2t_R-t) \ .
\label{fidelity}
\ee
At the instant $t=2t_R$
when the echo signal is usually acquired $r'(2t_R)=1$ and:
\ba
r(2t_R) & = & e^{i 2 E_{N'} t_R} \exp \left(- \frac { B_{N'}^2 (2 t_R)^2} 2 \right) \nonumber \\
& = & e^{i 2 \sum g_k' t_R} \exp \left( -2\sum (g_k')^2 t_R^2 \right) \ .
\label{signal}
\ea
Thus, reversal is incomplete. The deficit in the signal exhibits a Gaussian dependence on 
the instant of reversal $t_R$. This is the effect of the on-going decoherence 
due to ${\cal E}'$ -- these spins in the environment that did not get reversed. 

\begin{figure}
\centering \leavevmode
\epsfxsize 3.2in
\epsfbox{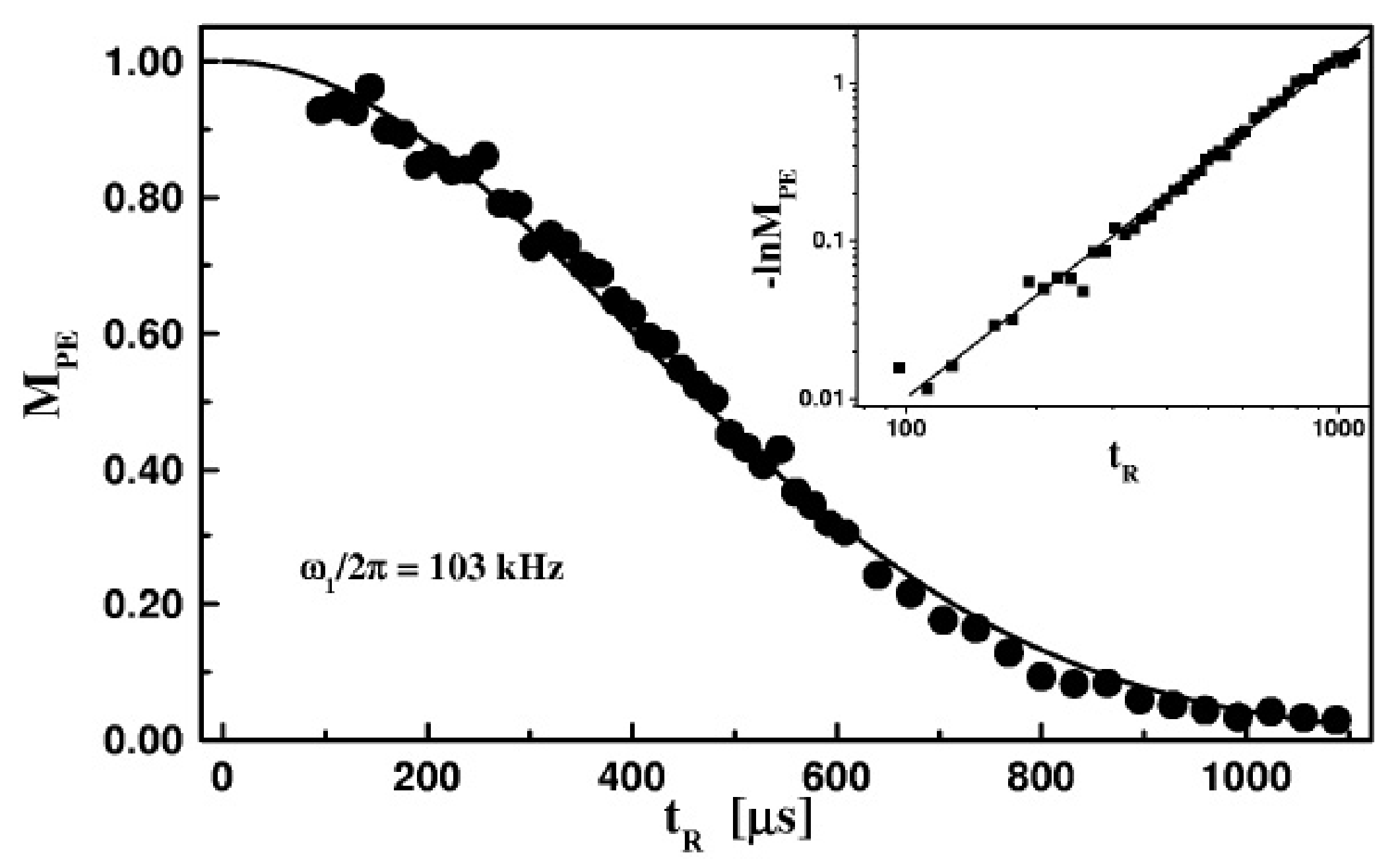}
\caption{Attenuation of the polarization echoes in a single crystal of ferrocene as a function of 
reversal time $t_R$: The data correspond to an orientation where the two molecules per unit cell 
are magnetically equivalent. The solid line corresponds to a Gaussian fitting yielding a characteristic 
time $T=(400\pm10)$ms as the single free parameter. The inset shows in effect $\ln (\mu (2t_0)-\mu(\infty))$
vs $t_0$ in a log-log plot.  The slope of the resulting line is $2.1\pm 0.08$, and is consistent 
with Eq. (\ref{signal})
(This illustration, Fig. 11 of Ref. \cite{horaciobook}, is reproduced here by permission of the authors.)}
\label{Figure4} 
\end{figure}

These equations exhibit the Gaussian time dependence (e.g., of the echo signal on the time
of reversal $t_R$) for large values of $t_R$ (i.e., beyond the initial quadratic regime) as was found 
in some of the Loschmidt echo experiments carried out by Levstein and Pastawski \cite{horaciobook} 
(see Fig. \ref{Figure4}). Most importantly, the partial reversal provides an explanation of the 
surprising experimentally observed {\it insensitivity} of the Gaussian decay of polarization to the 
details of the pulse that initiates reversal: As noted in Ref. \cite{horaciosaturation}, one might have expected that reversal pulse with larger amplitude will ``turn back'' evolution in a larger fraction
of the environment, but this does not seem to happen. Rather, independence of the ``backwards
evolution'' of the pulse inducing reversal indicates that always the same 
subset of the environment is turned back.
It is therefore tempting to interpret their experimental results using the ``two environment'' 
theory we have outlined above. 
We believe that such interpretation is basically correct, but that a more careful discussion 
should take into account differences between the system investigated in 
Ref. \cite{horaciobook} and our simple model.

\begin{figure}
\centering \leavevmode
\epsfxsize 2.0in
\epsfbox{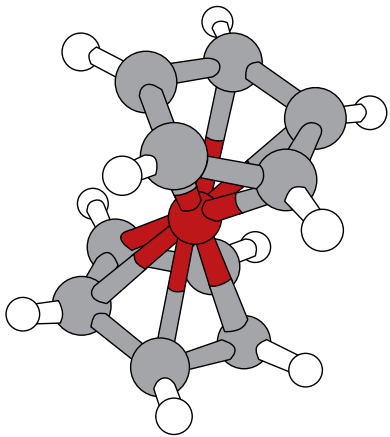}
\caption{Schematic representation of a ferrocene ($Fe(C_5H_5)_2$) molecule. The hydrogens
(white, on the outside) are coupled to the carbon atoms (light grey) which form a pentagon ring. 
The two rings are joined in the middle by an iron atom (dark grey). They can rotate with respect to 
each other and with respect to the solid matrix in which ferrocene is imbedded, which suggests
natural division into the immediate environment  that can be effectively reversed and the more distant
environment where the reversal is likely to fail.}
\label{FigFerrocene} 
\end{figure}

\begin{figure}
\centering \leavevmode
\epsfxsize 3.2in
\epsfbox{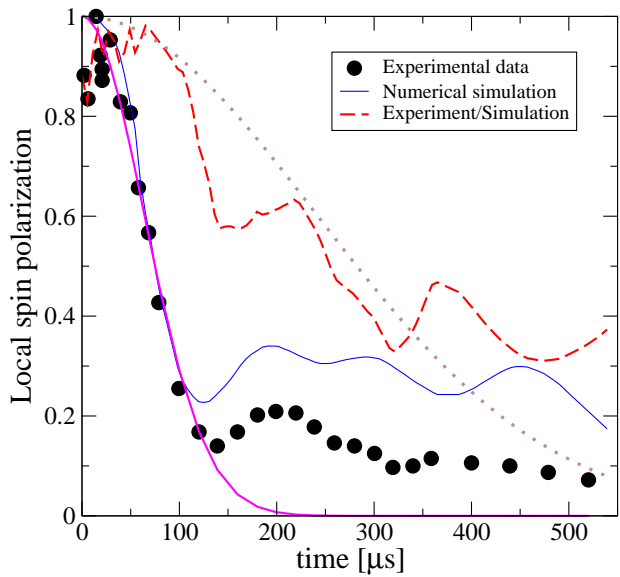}
\caption{Evolution of the local spin polarization in a single crystal of ferrocene.
The dots are the experimental data for a $0$ degree orientation of the crystal
with respect to the external magnetic field. The thin solid line is the calculated
evolution of the local polarization in a complete molecule where the rings rotate
independently in a staggered configuration. The dashed line is the ratio between
the simulation to the experimental data. The thick solid line is a Gaussian fit to the
initial experimental data (up to $100 \mu s$), and the thick dotted line is a fit to the
ratio between experiment and simulation. The characteristic decay times of the
fitted Gaussians are $90 \mu s$ and $340 \mu s$ respectively. This data is reproduced
from Fig. 8 of Ref. \cite{horaciobook} with permission from the authors.}
\label{LocalPol} 
\end{figure}

This view of the above data seems especially appropriate since in ferrocene ($Fe(C_5H_5)_2$,
the molecule used in Ref. \cite{horaciosaturation})
there are (at least) two environments that are likely to respond differently to the attempted 
``Loschmidt reversal'' of the dynamics. To point them out we need a bit more detailed
description of the experiment.
The ferrocene molecule (Fig. \ref{FigFerrocene}) consists of two rings, each with 5 hydrogen
atoms attached to 5 carbon atoms. The Loschmidt echo experiment starts when a rare $^{13}C$
atom (that appears in a small fraction of all the molecules) is polarized by the external field that starts 
the experiment. 
This polarization is then transferred to its adjacent hydrogen. Once it is there,
it can easily ``diffuse'' to the other hydrogens within the ring (or possibly within the molecule).
This process is rapid; a brief ($\sim 100 \mu s$) approximately Gaussian
decay leads to an ondulating plateau. The hydrogen adjacent to the $^{13}C$ atom is about $20\%$ 
polarized at this instant (see Fig. \ref{LocalPol}).

Up to that point, the agreement between the numerical simulation of quantum evolution in a single 
ferrocene molecule and the experiment is remarkable, suggesting that the only environment
explored by the injected polarization is the ``immediate neighborhood'': hydrogens within
the original $Fe(C_5H_5)_2$. Indeed, the value of the numerical plateau ($\sim \frac{1}{ 5}$ of the original
polarization) suggests that only the 5 hydrogens from the $C_5H_5$ ``ring'' that includes the rare
$^{13}C$ atom participate early on.

By contrast with this initial interval, there is a marked discrepancy between the experiment
and the single molecule simulations afterwards: Experimental data indicate leakage of the 
polarization from the molecule, with the signal decaying with time below the single molecule
numerical prediction (see Fig. \ref{LocalPol}). 
As time goes on, both the measured and the simulated polarizations
ondulate (indicating partial ``revivals'', presumably because of the finite size of the ferrocene 
molecule \cite{horaciomeso}) but the experimental data also indicates persistent polarization leakage. Over the same
time interval the simulation continues to hover just above $20\%$ of the original signal, and
exhibits at best only much slower systematic decay.

Given the previous discussion, we have now reached the ``eureka moment'': The immediate
environment -- hydrogens in the ferrocene (and, possibly, in only one of the $C_5H_5$ rings)
are responsible for short term approximately Gaussian decay, and for the partial revivals, 
consistent with the behavior of such small quantum systems (as seen in Fig. \ref{Figure1}). This is clearly
a good candidate for our ${\cal E}''$ -- the ``reversible'' part of the whole environment.

By contrast, once the signal leaks out to more distant ${\cal E}'$ (which is responsible
for the discrepancy between the single molecule simulations and the experimental data),
``the cat is out of the bag'', and it (e.g., the polarization which has leaked out of the molecule)
might be very difficult to recapture. This view is supported
by what is known about the structure of solid ferrocene: Individual molecules (and, indeed,
the two rings of the individual molecule) rotate on timescales short compared to 
these probed in the echo experiments. This dynamics will be much more difficult to reverse in the echo
experiment.

The reversal will then result in the desired echo only on subsystems in which
the atom has fixed neighbors (like the $C_5H_5$ ring), but is unlikely to succeed elsewhere.
So, the environment of the ferrocene molecule (neighboring ferrocene molecules, 
and possibly even its other ring) constitute ${\cal E}'$.

In closing this section we note that the timescale on which the echo decays is consistent
with 
a Gaussian fit to the experimental data of the decay of the local polarization
 divided by the numerical simulation of the isolated molecule, see Fig. \ref{LocalPol}.
The fit is consistent with an echo decay timescale of $4$ times the initial scale for decay
of the local polarization, i.e. $\sim 400 \mu s$ (see Fig. \ref{Figure4} for comparison).
This is an ecouraging observation. Further, this scale would not be affected by a better reversal
with the immediate environment ${\cal E}''$, consistent with the observed insensitivity
of the echo to radiofrequency power \cite{horaciosaturation}.

\section{Discussion} 

The model we have proposed is suggestive, but it is not yet conclusive: It offers only a rather
simplified representation of the experiment. For instance, it is much more reasonable to say that the polarization first diffuses to the immediate ``reversible'' environment, and that the more remote environment decoheres all of this reversible ${\cal E}''$ (and not just the original system). 
Nevertheless, the split into two environments -- our key assumption -- explains the key features of the 
data in a way that naturally fits the physics of the system.  However, it is useful to list at least
some of the approximations we have made, and to consider their implications.

To begin with, interactions between the spins in Ref. \cite{horaciobook} are dipolar, so the interaction 
Hamiltonian does not have the simple structure of the Ising Hamiltonian of Eq.~(\ref{hamiltonian0}).
Moreover, spins of the real environment interact with each other. Furthermore, interaction and self-Hamiltonians 
of the spins do not commute in general. 

Consequently, the straightforward manipulations that allowed us to derive Gaussian time dependence 
of the decoherence factor from first principles within a few lines cannot be directly carried out for 
more realistic models of the experiment. Nevertheless, the central ingredient needed to establish 
the Gaussian character of the echo does not seem to depend on these detailed assumptions. Rather, it is -- 
in essence -- the (approximately) Gaussian nature of the distribution of the eigenenergies of 
the total Hamiltonian, which then leads to the Gaussian time dependence of the decoherence factor. 
One can certainly believe that this very generic requirement is satisfied under conditions that are far 
more common than the specific assumptions of the simple decoherence model we have analyzed. 
Indeed, this broad applicability is the very essence of the central limit theorem we (and others \cite{HMH}) have invoked.

Even more convincing is the direct experimental evidence: Short time Gaussian dependence 
of the signal before reversal in the experiments involving ferrocene has been established \cite{horaciodifusion}
(see Fig. \ref{Figure4}). 
This is in effect the decoherence factor -- the characteristic function of the distribution of the relevant 
eigenenergies of the underlying Hamiltonian responsible for the evolution. And approximately 
Gaussian $r(t)$ implies (by the arguments involving Fourier transform) Gaussian eigenenergies. 

Time evolution of the NMR polarization signal is in such settings often interpreted as diffusion \cite{Abragam,DiffusionNMR}. 
This makes intuitive sense in the experiments that lead to Fig. \ref{Figure4}, as only rare nuclei of $C^{13}$ 
in a small fraction of ferrocene molecules are initially polarized, so the decay of the polarization signal 
is caused by the spreading of that polarization over an increasingly larger environment. However,
this effective diffusion must obviously reflect a reversible dynamical process generated by an underlying 
Hamiltonian, as fundamentally diffusive evolution could never be reversed.
This is reflected in the short time mesoscopic echoes observed in this ``diffusive'' process \cite{horaciomeso}
due to the small size of the first environment.
To account for 
the diffusive character of the evolution the distribution of eigenenergies, $\eta(E)$ must be Gaussian 
in character. So, while specific assumptions we used in our simple model are not satisfied in the experimental setting, Gaussianity of the energy spectrum we were led to as a result of these assumptions may well turn out to be a fairly generic feature. 

\section{Summary and Conclusions}

We have seen how -- in the quantum setting of decoherence and Loschmidt echo -- deterministic 
dynamics can lead to evolutions that have a distinctly stochastic Gaussian character. While our model 
is rather simple and clearly too idealized to directly address experimental situation of Refs. 
\cite{horaciobook,horacioexponential,horaciodifusion,horaciosaturation}, 
it also suggests that our main results -- Gaussian decay of the decoherence factor and Gaussian echo --
will appear whenever the energy spectrum of the excitation corresponding to the initial state of
the system is approximately Gaussian. As we have noted earlier, 
there is ample evidence of this in the existing experiments
involving ferrocene. Even more, we can reproduce the observed 
insensitivity to perturbations of the Gaussian echo decay. Qualitative -- and even quantitative --
 comparisons between predictions of our model and the experimental data are promising. 




We acknowledge useful discussions with Horacio M. Pastawski and Patricia R. Levstein. One of the authors (WHZ) acknowledges the Alexander von Humboldt Foundation Prize and the hospitality of the Center for Theoretical Physics of the University of Heidelberg, where part of this work was carried out.

\end{document}